\documentclass[iop]{emulateapj}

\shorttitle{POSSIBLE WATER IN AN EXOPLANETARY BODY}
\shortauthors{Farihi et al.}

\slugcomment{Accepted to ApJ Letters}

\begin{document}

\title{POSSIBLE SIGNS OF WATER AND DIFFERENTIATION IN A ROCKY EXOPLANETARY BODY}

\author{J. Farihi\altaffilmark{1},
	C. Brinkworth\altaffilmark{2,3},
	B. T. G\"ansicke\altaffilmark{4},
	T. R. Marsh\altaffilmark{4},
	J. Girven\altaffilmark{2,4},
	D. W. Hoard\altaffilmark{2},
	B. Klein\altaffilmark{5},
	\& D. Koester\altaffilmark{6}}

\altaffiltext{1}{Department of Physics \& Astronomy,
			University of Leicester,
			Leicester LE1 7RH, UK; 
			jf123@star.le.ac.uk}
			
\altaffiltext{2}{Spitzer Science Center,
			California Institute of Technology,
			MS 220-6,
			Pasadena, CA 91125}

\altaffiltext{3}{NASA Exoplanet Science Institute, 
			California Institute of Technology, 
			Pasadena, CA 91125}		

\altaffiltext{4}{Department of Physics, 
			University of Warwick, 
			Coventry CV5 7L, UK}	
			
\altaffiltext{5}{Department of Physics \& Astronomy, 
			University of California, 
			Los Angeles, CA 90095}
			
\altaffiltext{6}{Institut f\"ur Theoretische Physik und Astrophysik, 
			University of Kiel, 
			24098 Kiel, Germany}

\begin{abstract}

{\em Spitzer} observations reveal the presence of warm debris from a tidally destroyed rocky and 
possibly icy planetary body orbiting the white dwarf GD\,61.  Ultraviolet and optical spectroscopy of 
the metal-contaminated stellar photosphere reveal traces of hydrogen, oxygen, magnesium, silicon, 
iron, and calcium.  The nominal ratios of these elements indicate an excess of oxygen relative to that 
expected from rock-forming metal oxides, and thus it is possible that water was accreted together 
with the terrestrial-like debris.  Iron is found to be deficient relative to magnesium and silicon, 
suggesting the material may have originated as the outer layers of a differentiated parent body, 
as is widely accepted for the Moon.

\end{abstract}


\section{INTRODUCTION}

In the search for terrestrial planetary systems around other stars, white dwarfs offer a unique
astrophysical advantage.  Owing to high surface gravities and the onset of convection, any 
atmospheric metals sink rapidly as these Earth-sized stellar embers cool below 25\,000\,K, 
leaving behind only H or He in the outermost layers of the star \citep{koe09}, a physical 
process corroborated by observation \citep{koe05,zuc03}.  Those stars with rocky planetary 
system remnants can become contaminated by the accretion of small, but spectroscopically 
detectable, amounts of heavy elements.  Metal lines in cool white dwarfs are a telltale of external 
pollution that typically implies either ongoing metal accretion rates $\dot{M_{\rm z}} >10^8$\,g\,s$
^{-1}$ \citep{koe06}, or asteroid-sized masses of heavy elements within the convection zone of the 
star \citep{far10a}.

To date, metal-rich dust and gas disks \citep{jur09a,gan08,gan06,rea05}, very likely produced by 
the tidal disruption of large asteroids \citep{deb02}, have been found closely orbiting more than 
one dozen cool white dwarfs \citep{far09,von07,jur07} and provide a ready explanation for the 
metal absorption features seen in their atmospheres \citep{jur03}.  The circumstellar material being 
gradually accreted by the white dwarf can be directly observed in the stellar photosphere to reveal 
its elemental abundances \citep{kle10,zuc07}.  While transit spectroscopy of main-sequence stars 
offers some information on the chemical composition of extrasolar gas giant atmospheres, metal-$
$polluted white dwarfs distill entire rocky planetary bodies into their constituent elements.  

This letter reports the detection of circumstellar dust at the white dwarf GD\,61, a He-rich 
star whose atmosphere is highly polluted with O, Si and Mg, and to a lesser degree by Fe and 
Ca.  With the exception of Mg detected in this work, these photospheric heavy elements were 
identified previously via far-ultraviolet \citep{des08} and optical \citep{sio88} spectroscopy, but 
interpreted as products of convective dredge-up and interstellar accretion.  The analysis here 
re-interprets these elements as the result of infall from closely orbiting circumstellar debris that 
was once contained in a sizable planetary body such as a large asteroid.

\section{DATA AND ANALYSIS}

\subsection{Circumstellar Dust}

\begin{figure}[hb!]
\epsscale{1.1}
\plotone{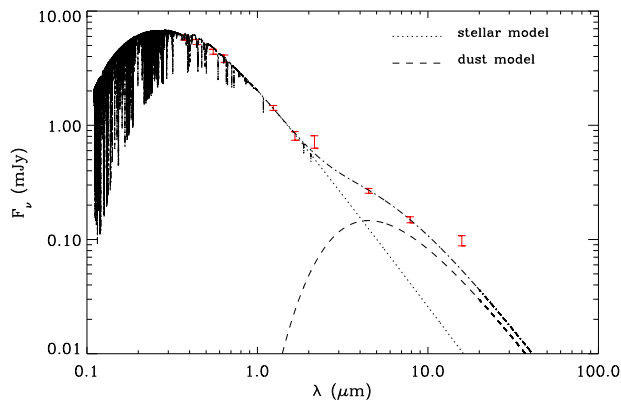}
\caption{Ultraviolet through infrared energy distribution of GD\,61 with photometry represented as 
error bars.  Fluxes below 2\,$\mu$m (the 2MASS $K_s$-band measure appears inaccurate and
has been ignored) are fitted with a model atmosphere, yielding a photometric distance of 50.1\,pc.  
The dashed line is a flat disk model with inner and outer temperatures of 1300 and 1000\,K at 
$i = 79^{\circ}$.  The IRS blue peak-up flux is likely to contain significant silicate emission.
\label{fig1}}
\end{figure}

\begin{deluxetable*}{rllcc}
\tabletypesize{\footnotesize}
\tablecaption{Observed and Modeled Properties of GD\,61\label{tbl1}}
\tablewidth{0pt}
\tablehead{
\multicolumn{2}{c}{Parameters}		&
\multicolumn{3}{c}{Fluxes}\\
\\
\colhead{}					&
\colhead{}					&
\colhead{Filter}				&
\colhead{$\lambda_{\rm eff}$ ($\mu$m)}	&
\colhead{$F_{\nu}$ (mJy)}}

\startdata

$T_{\rm eff}$					&17\,280\,K					&$U$		&0.37		&$5.87\pm0.27$\\
$\log\,g$\,(cm\,s$^{-2}$)			&8.20						&$B$		&0.44		&$5.36\pm0.27$\\
$d$							&50.1\,pc						&$V$		&0.55		&$4.41\pm0.22$\\
$M$							&0.71\,$M_{\odot}$				&$r'$			&0.64		&$3.84\pm0.28$\\
$M_{\rm ms}$					&3.0\,$M_{\odot}$ (A0 V)			&$J$			&1.24		&$1.42\pm0.07$\\
Total Age						&0.6\,Gyr						&$H$		&1.66		&$0.81\pm0.07$\\
$(\mu_{\alpha}$,$\mu_{\delta})$	&($-19,-102$)\,mas\,yr$^{-1}$		&$K_s$		&2.16		&$0.72\pm0.09$\\
$z_{\rm g}c$					&$+$41\,km\,s$^{-1}$			&IRAC\,2		&4.49		&$0.267\pm0.013$\\
$v_{\rm rad}$					&$+$17\,km\,s$^{-1}$			&IRAC\,4		&7.87		&$0.149\pm0.009$\\
$(U,V,W)$						&($10,-4,-13$)\,km\,s$^{-1}$		&IRS\,Blue	&15.8		&$0.098\pm0.010$

\enddata

\tablecomments{Temperature and surface gravity are from \citet{des08} and provide a good fit to 
the HIRES spectrum.  The total age estimate is the sum of the white dwarf cooling age \citep{fon01} 
and main-sequence lifetime \citep{hur00}.  The radial velocity is the difference between the average 
measured line velocity of $+58$\,km\,s$^{-1}$ and the gravitational redshift of the star.  Short wavelength 
photometry and proper motion are taken from the literature and available catalogs.}

\end{deluxetable*}

{\em Spitzer Space Telescope} \citep{wer04} observations of GD\,61 were executed on 2009 
March 18 and 7 April during the telescope's final cryogenic cycle.  Imaging photometry was 
performed with the Infrared Array Camera (IRAC; \citealt{faz04}) at 4.5 and 7.9\,$\mu$m, and with 
the Infrared Spectrograph (IRS; \citealt{hou04}) in the blue peak-up filter.  The data were reduced 
and photometry executed following \citet{far09} and \citet{bri09}; results are listed in Table \ref{fig1} 
and plotted in Figure \ref{fig1}.

The flux densities measured with {\em Spitzer} exhibit an infrared excess, consistent with warm 
dust orbiting within the tidal breakup (Roche) limit of the star \citep{von07}.  These data cannot
be reproduced with a low-mass stellar or brown dwarf secondary \citep{pat06,far05}.  In Figure 
\ref{fig1}, a model stellar atmosphere is fitted to the ground-based photometry, and the addition of 
an optically thick disk model with a flat geometry is shown to reproduce the IRAC data rather well.  
The measured 15.8\,$\mu$m IRS peak-up flux is likely to contain silicate emission, as observed in 
G29-38 \citep{rea09}, and strongly suspected in SDSS\,1228 \citep{bri09}.  

The temperatures of the disk model correspond to inner and outer disk edges of 19 and 26 stellar
radii (0.21 and 0.30\,$R_{\odot}$), respectively.  Thus, the dust ring is relatively narrow ($\Delta r < 0.1
\,R_{\odot} < r_{\rm inner}$ ) as found for a growing number of metal-polluted white dwarfs \citep{mel11,
far10b}.  Otherwise, the infrared continuum at GD\,61 is fairly typical of white dwarf disks studied with 
{\em Spitzer} \citep{far09,jur07}, and consistent with closely orbiting debris resulting from the tidal 
destruction of a planetary body such as a large asteroid.

\subsection{Atmospheric Heavy Elements}

\begin{figure*}
\epsscale{1.1}
\plotone{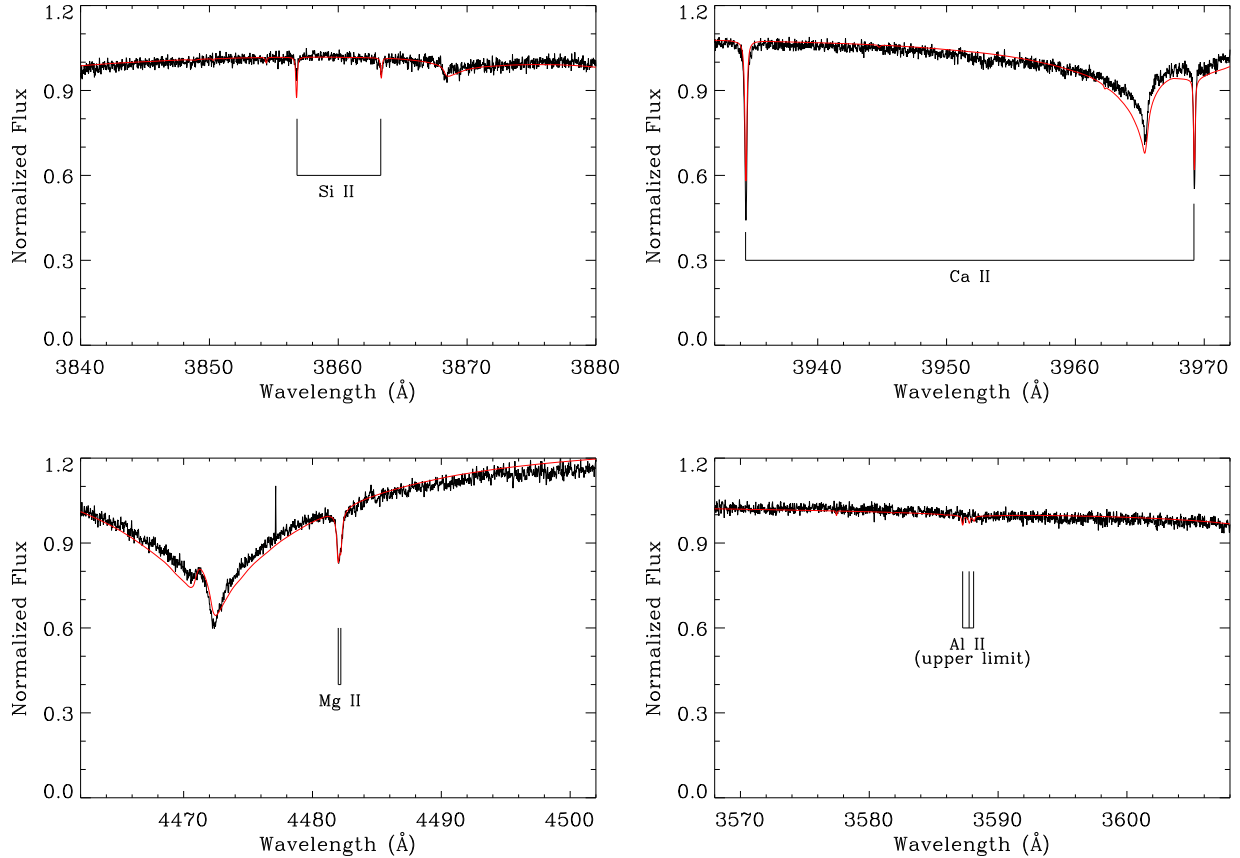}
\caption{HIRES data for GD\,61, showing regions containing key element detections and upper 
limits.  The unsmoothed data are shown in black and the model overplotted in red.  There are 
unavoidable and slight imperfections in the flux calibration, and these can be seen relative to 
the model; however, these do not affect the abundance determinations.
\label{fig2}}
\end{figure*}

GD\,61 was observed using the High-Resolution Echelle Spectrograph (HIRES; \citealt{vog94}) 
at the Keck I telescope on Mauna Kea.  Spectra covering $3130-5960$\AA \ (with gaps at $4004-4
064$, $4992-5071$\AA) were acquired on 2008 February 13, and the $R\sim40\,000$ data were 
reduced and analyzed following \citet{kle10}.  Heavy elements detected in the HIRES spectrum are: 
\ion{Mg}{1} (3838\AA); \ion{Mg}{2} (4481\AA); \ion{Si}{2} (3586, 3863, 4128, 4131\AA); and \ion{Ca}{2} 
(3159, 3179, 3934, 3969\AA).  Figure \ref{fig2} displays a few important sections of the spectrum.

Element abundances were derived from comparisons of equivalent widths in the data with those 
of stellar atmosphere models, with upper limits established from the absence of model-predicted 
features.  The average Ca abundance was weighted by feature strength, giving most weight to the 
well-studied \ion{Ca}{2} H \& K resonance lines.  For Mg and Si, the abundances for the sets of lines 
were combined in a direct average.  The uncertainties have been calculated in a manner similar to 
that described in \citet{kle10}, by varying the model temperature and surface gravity by $\pm400$\,K 
and $\pm0.3$\,dex, respectively.

Prior observations of GD\,61 with the {\em Far Ultraviolet Spectroscopic Explorer (FUSE)}
were obtained and analyzed in detail by \citet{des08}.  In order of decreasing abundance, these 
data reveal trace amounts of H, O, Si, and Fe in the He-dominated atmosphere of the white dwarf.  
Modeled stellar parameters found by \citet{des08} are listed in Table \ref{tbl1} and were found to fit 
the He lines in the HIRES spectrum of GD\,61 quite well overall.  Furthermore, these parameters are 
consistent with the ultraviolet through infrared photometric energy distribution of the star shown in 
Figure \ref{fig1}, and were adopted for all purposes in this work, including the determination of 
metal abundances and upper limits from the HIRES dataset.

Only Si is detected in both the {\em FUSE} and HIRES datasets, and the independent analyses 
agree rather well, as shown in Table \ref{tbl2}.  Also, the upper limit Fe abundance determined from 
the HIRES spectrum is consistent with its reported {\em FUSE} detection.  In order to preserve the 
metal-to-metal ratios found for each spectrum independently, all heavy element ratios were tied 
to the weighted average of the {\em FUSE} and HIRES Si abundances.  Table \ref{tbl2} lists the
HIRES, {\em FUSE}, and adopted heavy element abundances relative to He, while carrying the 
errors distinct to each dataset.

\subsection{Infalling Debris from a Destroyed Minor Planet}

Rather notably, O is the most abundant heavy element in GD\,61, whereas only a strict upper 
limit for C could be determined from the ultraviolet data.  Lacking information on its circumstellar
environment, \citet{des08} had to invoke convective dredge-up combined with a highly unusual 
chemical profile of the white dwarf core to account for the extreme O/C $>1000$ found in GD\,61.
However, all white dwarfs suspected of harboring dredged-up core material exhibit atmospheric C 
and effective temperatures near or below 12\,000\,K \citep{duf05}, including the recently detected 
class of O-rich white dwarfs \citep{gan10}.  At 17\,300\,K, the depth of the convective envelope in 
GD\,61 is only 2.5\% of the maximum depth attained in He atmosphere white dwarfs \citep{koe09}, 
which occurs between 12\,000 and 10\,000K \citep{pel86}.  Furthermore, dredge-up cannot account 
for detectable amounts of Si, Ca, and Fe in the atmosphere of a white dwarf, and \citet{des08} 
concluded that GD\,61 must have accreted these metals from the interstellar medium.  However, 
they did not consider that metals are locked up in dust grains within the interstellar medium, and 
this dust is composed of O-rich silicates \citep{dra03}.  Such material provides a natural source 
for the observed O if interstellar silicates were accreted.

The warm dust orbiting metal-rich white dwarfs and mature stars in general is rich in silicate 
minerals \citep{jur09a,lis08}, and hence a large deposition of Si and O should result from the 
accretion of this debris, in excellent qualitative agreement with the polluted atmosphere of GD\,61.
Accretion from the closely orbiting dust disk must be ongoing due to viscous dissipation; primarily 
gas drag at the inner disk edge where dust grains rapidly sublimate \citep{jur08}.  Material found 
orbiting within 20 stellar radii demonstrates that the white dwarf is currently accreting from this 
reservoir and provides a natural explanation for all the photospheric heavy elements, including O.  

\section{RESULTS}

\subsection{Chemical Abundances of the Debris}

\begin{deluxetable*}{ccccccc}[h]
\tabletypesize{\footnotesize}
\tablecaption{Heavy Elements in the Disrupted Minor Planet Orbiting GD\,61\label{tbl2}}
\tablewidth{0pt}
\tablehead{
\colhead{}								&
\colhead{}								&
\colhead{}								&
\colhead{$\log\,[n({\rm Z})/n({\rm He})]$}		&
\colhead{}								&
\colhead{Early Phase}					&
\colhead{Steady State}\\
\colhead{Element}						&
\colhead{$t_{\rm diff}$ ($10^5$\,yr)}			&
\colhead{HIRES}						&
\colhead{{\em FUSE}}					&
\colhead{Adopted}						&
\colhead{$M_{\rm z}$ (10$^{21}$\,g)}		&
\colhead{$M_{\rm z}$ (10$^{21}$\,g)}}

\startdata

H		&$\infty$		&$-3.98\pm0.02$		&\nodata			&$-3.98\pm0.02$		&\phm{....}5.615	&5.615\\
C		&1.245		&$<-8.8$				&\nodata			&$<-8.93$		 	&$<0.001$		&\nodata\\
O		&0.938		&\nodata				&$-5.8\pm0.2$		&$-5.93\pm0.20$		&\phm{....}1.017	&0.873\\
Mg		&0.794		&$-6.65\pm0.18$		&\nodata			&$-6.63\pm0.18$		&\phm{....}0.308	&0.312\\
Al		&0.685		&$<-7.2$				&\nodata			&$<-7.18$		 	&$<0.097$		&\nodata\\
Si		&0.650		&$-6.85\pm0.09$		&$-6.7\pm0.2$		&$-6.83\pm0.08$		&\phm{....}0.225	&0.278\\
Ca		&0.501		&$-7.90\pm0.19$		&\nodata			&$-7.88\pm0.19$		&\phm{....}0.029	&0.046\\
Ti		&0.436		&$<-9.1$				&\nodata			&$<-9.08$		 	&$<0.002$		&\nodata\\
Cr		&0.396		&$<-9.0$				&\nodata			&$<-8.98$		 	&$<0.003$		&\nodata\\
Fe		&0.361		&$<-7.5$				&$-7.6\pm0.2$		&$-7.73\pm0.20$		&\phm{....}0.056	&0.125\\
Ni		&0.346		&$<-7.6$				&\nodata			&$<-7.58$			&$<0.084$		&\nodata
	
\enddata

\tablecomments{Adopted abundances, upper limits and errors are discussed in \S2.  The last two 
columns list the cumulative mass of each element residing in the convection zone of GD\,61 for the 
early phase and steady state cases.  Thus, the accreted metals in GD\,61 total a minimum of $1.64
\times10^{21}$\,g, while an ongoing event older than $2.5\times10^5$\,yr implies a total accreted
mass exceeding 10$^{23}$\,g.}

\end{deluxetable*}

The heavy elements observed in GD\,61 represent the chemical constituency of its circumstellar 
disk, and their abundance ratios provide constraints on the nature of the minor planet pulverized
into the orbiting debris.  These elements reside in the $2.15\times10^{26}$\,g stellar convection 
zone, where they are thoroughly mixed and have sinking timescales on the order of $10^{4.5}$\,yr 
\citep{koe09}.  Table \ref{tbl2} lists the current (early phase) abundances and masses of each heavy 
element detected in the star, as well as upper limits established from the HIRES and {\em FUSE} 
spectra.  Based on these values, and because metals continually sink below the photosphere, the 
destroyed planetary body had a {\em minimum} mass of $1.64\times10^{21}$\,g, roughly equivalent 
to an asteroid 110\,km in diameter.

If the disk age is less than around 10$^3$\,yr ($\ll t_{\rm diff}$), then insufficient time has passed 
for any metal to substantially diffuse below the photosphere.  Observable signatures of the element 
ratios and the total accreted mass of the disrupted asteroid are preserved in the star during this early 
phase of accretion.  On the other hand, if the disk age approaches or exceeds several $t_{\rm diff}$, a 
steady state balance between accretion and diffusion is achieved for each heavy element \citep{koe06}.
In either case, the abundance ratios and minimum element masses in the minor planet can be derived 
analytically, without knowing the exact age of the disk, and are listed in Table \ref{tbl2}.

\subsubsection{Potential Evidence of H$_2$O}

Both the early phase and steady state scenarios predict that the total mass of observed heavy
elements in GD\,61 is dominated by O.  If all the Mg, Si, Ca, and Fe were originally contained in 
MgO, SiO$_2$, CaO, and FeO, as in chondrites and the rocky material of the inner Solar System,
the nominal {\em FUSE} and HIRES abundances indicate a significant O excess.  Table \ref{tbl3} 
explores the fraction of O contained in these metal oxides, together with upper limit and chondritic 
Al delivered as Al$_2$O$_3$.  Even in the case where Al is present at its upper limit abundance 
as determined by HIRES, there is a nominal excess mass of O by 22\% or 44\% in the steady state 
or early phases, respectively.  However, at the minimum values of O/Mg and O/Si permitted by the 
errors, the O can be accounted for completely by dry mineral oxides, and thus an excess cannot 
be confirmed with confidence using the present data.

The potential for excess O in GD\,61 is of interest because the most natural explanation would be 
that extra O was originally bound in water ice representing a significant fraction of the total mass of 
the minor planet; 25\% to 35\% in the steady state.  It is possible for water to survive the post-main 
sequence evolution of the star, if sufficiently buried within its parent body \citep{jur10}.  Upon 
shattering during a close approach with a white dwarf, any water ice (and volatiles) would rapidly 
sublimate but eventually fall onto the star -- the feeble luminosity of white dwarfs is insufficient to 
remove even light gases by radiation pressure \citep{far08}.

If the system is still in an early phase of accretion, the potential for water is enhanced, as the implied
O abundance of the accreted material is then at its highest.  Given that disks at white dwarfs appear 
to be long-lived -- and in this way potentially analogous to planetary rings in the Solar System -- with 
lifetimes possibly exceeding $10^6$\,yr \citep{kle10,jur09b,jur08,far08}, the chance of catching the 
star in the early phase is small.  Regardless, the primary source of uncertainty for a water-rich accretion 
event at GD\,61 is the abundance errors on O, Mg, and Si.

\subsubsection{Possible Evidence of Differentiation}

GD\,61 is polluted by material that is deficient in Fe relative to both Mg and Si.  Chondritic and 
solar ratios among these elements are near unity; Fe/Mg $=0.83$, Fe/Si $=0.85$ \citep{lod03}.  
These same ratios in GD\,61 are 0.08, 0.13 in the early phase, and 0.18, 0.23 in the steady state, 
respectively.  In either case, the accreted material was depleted in Fe relative to the material of 
the inner Solar System.

If the remnant planetary system at GD\,61 formed at roughly chondritic element ratios, then 
some process must have depleted Fe from the parent body that now exists as a circumstellar ring 
of debris.  Exactly such an occurrence is the favored hypothesis for the origin of the Moon.  Near
the end of terrestrial planet formation, a Mars-sized body impacted the Earth, ejecting the mantle
of the impactor while the cores of the two bodies liquified and fused \citep{can01}.  If the event
happened sufficiently late that both bodies were largely differentiated, the ejected mantle would
be relatively Fe-poor, giving rise to the observed deficiency in the Moon.

It is unknown if extrasolar terrestrial bodies form at chondritic compositions or if significant 
variations exist.  Studying destroyed rocky planetary bodies at white dwarfs will help shed light
on this important and interesting question.  It is conceivable that the minor (or major) planet that 
gave rise to the metals in GD\,61 was formed with a distinctly non-solar Fe abundance.

\begin{deluxetable}{ccccc}
\tabletypesize{\footnotesize}
\tablecaption{Assessment of Oxygen Mass Fractions in GD\,61\label{tbl3}}
\tablewidth{0pt}
\tablehead{							&
\multicolumn{2}{c}{Early Phase}	&
\multicolumn{2}{c}{Steady State}\\
\\
\colhead{Oxygen Carrier}			&
\colhead{Fraction}				&
\colhead{Error\tablenotemark{a}}	&
\colhead{Fraction}				&
\colhead{Error\tablenotemark{a}}}

\startdata

SiO$_2$					&\phm{....}0.252	&0.110			&\phm{....}0.363	&0.158\\
MgO						&\phm{....}0.200	&0.115			&\phm{....}0.236	&0.136\\\
Al$_2$O$_3$			  	&$<0.084$		&\nodata	 		&$<0.115$		&\nodata\\
FeO\tablenotemark{b}		&\phm{....}0.016	&0.010			&\phm{....}0.041	&0.025\\
CaO						&\phm{....}0.011	&0.007			&\phm{....}0.021	&0.013\\
\\

\hline
\\
Excess (Chondritic Al)		&\phm{....}0.498	&0.223			&\phm{....}0.307	&0.304\\
Excess (Maximum Al)		&\phm{....}0.437	&0.246			&\phm{....}0.224	&0.334

\enddata

\tablenotetext{a}{Standard deviations resulting from all possible permutations of individual element 
abundances and associated errors.}

\tablenotetext{b}{The mass fraction of Fe is insufficient for the outcome to be affected by significant 
amounts of Fe$_2$O$_3$, which is only present in significant quantities in some Earth basalts, but 
not in the bulk of meteorites, including Lunar meteorites and Eucrites \citep{bas81}.}

\end{deluxetable}

\section{OUTLOOK}

The present uncertainties in the O/Mg and O/Si ratios in GD\,61 permit an interpretation where 
the accretion of water is not required to account for its O-rich atmosphere.  In order to reduce these 
errors, a spectrum containing all the Table \ref{tbl2} detected elements is needed, whereby metal 
abundances can be tied to O (rather than He) and a confident assessment of its oxide contribution 
can be made.  

If the excess O can be confirmed with better data, the accretion of a water-rich asteroid is the 
most natural explanation for the circumstellar debris detected by {\em Spitzer} and the pattern of 
elements in the disk-polluted star.  If all the trace H in GD\,61 was delivered by a single asteroid 
that has polluted the star for many diffusion timescales (metals continuously sink, while H floats),
then the parent body contained $5.0\times10^{22}$\,g of water.  If such a minor planet were 25\% to 
35\% water by mass as inferred from the steady state analysis and nominal O abundance, then its 
total mass would be 1.4 to $2.0\times10^{23}$\,g and approaching that of Vesta, the second most
massive asteroid in the Solar System.  Such a large mass of orbiting material is permitted by the 
optically thick disk model shown in Figure \ref{fig1} \citep{jur03}, while the steady state scenario
implies a comparable mass has already been accreted by the star.

GD\,61 appears to have a relatively young total age.  This picture is supported by the 
spectroscopically determined mass and also by its three dimensional space velocity vector, 
$(U,V,W)$ in Table \ref{tbl1}.  Its current mass of 0.71\,$M_{\odot}$ implies a likely descent from a 
main-sequence A0 V star of around 3.0\,$M_{\odot}$ \citep{wil09,kal08,dob06}, and its overall mild 
velocity is consistent with relatively young, thin disk kinematics.  A total age of 0.6\,Gyr is inferred 
from its main-sequence lifetime plus cooling age \citep{fon01,hur00}.  Thus, the debris orbiting 
GD\,61 is material associated with a rocky planetary system formed about a relatively young, 
intermediate-mass star. 

Recently, significant amounts of water have potentially been identified in the asteroid belt 
of the Solar System, via main belt comet activity \citep{hsi06} and surface water on Themis 
\citep{cam10,riv10}.  These discoveries have important implications for the delivery of water to 
a dry Earth, a task that appears most readily accomplished by primordial asteroids -- planetary 
embryos analogous to Ceres -- in the outer main belt \citep{mor00}.  Ceres itself is thought to 
harbor significant water ice in its interior; its density and dynamical modeling suggest a fraction
of 25\% water by mass \citep{tho05}.  The white dwarf GD\,61 is surrounded by a ring of dust, 
and the totality of its atmospheric pollution can be understood as the tidal destruction of a parent 
body similar in mass to Vesta.  If confirmed, the potential O excess would suggest a water content
analogous to Ceres.  These two largest Solar System asteroids are (just) massive enough to be 
differentiated, and the metals in GD\,61 suggest its circumstellar debris may have originated 
in a similarly massive, rocky exoplanetary body.

\acknowledgments

The authors thank the anonymous referee for a careful review and helpful comments.  This work 
is based in part on observations made with the {\em Spitzer Space Telescope}, which is operated 
by the Jet Propulsion Laboratory, California Institute of Technology under a contract with NASA.  
Some of the data presented herein were obtained at the W. M. Keck Observatory, which is 
operated as a scientific partnership among the California Institute of Technology, the University 
of California and NASA.  This publication makes use of data products from the Two Micron All Sky 
Survey.

{\em Facility:} \facility{{\em Spitzer} (IRAC)}; \facility{{\em Keck} (HIRES)}

\end{document}